\documentclass[a4paper,twocolumn,superscriptaddress,accepted=2025-05-27]{quantumarticle}
\pdfoutput=1

\usepackage[utf8]{inputenc}
\usepackage[american]{babel}
\usepackage[T1]{fontenc}
\usepackage[pdftex]{graphicx}  
\usepackage{xcolor}
\usepackage{dcolumn}
\usepackage{bm}
\usepackage{dsfont}
\usepackage{amsmath,amsthm,amssymb}
\usepackage{braket}
\usepackage{wrapfig}
\usepackage{mathtools}
\usepackage{float}
\usepackage{physics}
\usepackage[normalem]{ulem}

\usepackage{hyperref}
\hypersetup{colorlinks,bookmarksopen,bookmarksnumbered,citecolor=red,linkcolor=red,pdfstartview=false,urlcolor=red}
\usepackage[numbers,sort&compress]{natbib}
\newcommand{\daga}{^\dag}

\begin{document}
\title{Breakdown of Measurement-Induced Phase Transitions Under Information Loss}

\author{Alessio Paviglianiti}\thanks{Equal contribution}
\affiliation{International School for Advanced Studies (SISSA), via Bonomea 265, 34136 Trieste, Italy}
\author{Giovanni Di Fresco}\thanks{Equal contribution}
\affiliation{Dipartimento di Fisica e Chimica ``Emilio Segr\`e'', Group of Interdisciplinary Theoretical Physics, Universit\`a degli Studi di Palermo, Viale delle Scienze, Ed. 18, I-90128
Palermo, Italy}
\author{Alessandro Silva}
\affiliation{International School for Advanced Studies (SISSA), via Bonomea 265, 34136 Trieste, Italy}
\author{Bernardo Spagnolo}
\affiliation{Dipartimento di Fisica e Chimica ``Emilio Segr\`e'', Group of Interdisciplinary Theoretical Physics, Universit\`a degli Studi di Palermo, Viale delle Scienze, Ed. 18, I-90128
Palermo, Italy}
\affiliation{Stochastic Multistable Systems Laboratory, Lobachevsky University, 603950 Nizhniy Novgorod, Russia}
\author{Davide Valenti}
\affiliation{Dipartimento di Fisica e Chimica ``Emilio Segr\`e'', Group of Interdisciplinary Theoretical Physics, Universit\`a degli Studi di Palermo, Viale delle Scienze, Ed. 18, I-90128
Palermo, Italy}
\author{Angelo Carollo}
\affiliation{Dipartimento di Fisica e Chimica ``Emilio Segr\`e'', Group of Interdisciplinary Theoretical Physics, Universit\`a degli Studi di Palermo, Viale delle Scienze, Ed. 18, I-90128
Palermo, Italy}

\begin{abstract}
   The dynamics of a quantum-many body system subject to measurements is naturally described by an ensemble of quantum trajectories, which can feature measurement-induced phase transitions (MIPTs). This phenomenon cannot be revealed through ensemble-averaged observables, but it requires the ability to discriminate each trajectory separately, making its experimental observation extremely challenging. We explore the fate of MIPTs under an observer’s reduced ability to discriminate each measurement outcome. This introduces uncertainty in the state of the system, causing observables to probe a restricted subset of trajectories rather than a single one. By introducing an exactly-solvable Liouvillian model, we examine how long-time spatial correlations are influenced by varying degrees of trajectory averaging. We compute exactly the correlation matrix, Liouvillian gap, and entanglement negativity to demonstrate that averaging over multiple realizations introduces an effective finite lengthscale, beyond which long-range correlations are suppressed. This suggests that partial averaging over trajectories conceals the critical features of individual realizations, thereby blurring away the signatures of distinct measurement-induced phases.
\end{abstract}
\maketitle

\section{Introduction}
Understanding and classifying the dynamical behavior of many-body quantum systems remains a long-standing and complex challenge in condensed matter physics~\cite{polkovnikov2011,heyl2018}. Out-of-equilibrium closed systems give rise to an extremely diverse phenomenology in terms of thermalization~\cite{deutsch1991,rigol2008,dalessio2016,gring2012,mori2018,calabrese2016,gogolin2016,vidmar2016,ueda2020}, transport~\cite{heidrichmeisner2007,vasseur2016,deluca2013,ljubotina2017,ye2020}, and entanglement properties~\cite{amico2008,calabrese2005,bardarson2012,nahum2017,laflorencie2016,pappalardi2018,alba2018}. Recently, the question of how these features are influenced when the time evolution is perturbed by measurements has garnered increasing interest. This attention mostly stems from measurement-induced phase transitions (MIPTs), where the interplay between unitary dynamics and external monitoring creates distinct non-equilibrium phases characterized by different entanglement structures~\cite{li2018,skinner2019,gullans2020,koh2023,cao2019,choi2020,poboiko2023,block2022,minato2022,zabalo2020,li2019,sharma2022,bao2020,turkeshi2021,paviglianiti2023,piccitto2022,barratt2022,coppola2022,fux2023,passarelli2024}. Since their discovery, MIPTs have been extensively explored in various contexts, ranging from quantum circuits to Hamiltonian models.

Due to the intrinsic stochastic nature of quantum measurements, the time evolution of a system undergoing external monitoring is inherently non-deterministic. Different realizations of the same protocol yield different quantum trajectories~\cite{wiseman1996,jacobs2006,brun2002,daley2014}, each specified by a unique collection of measurement outcomes, resulting in different final states. Experimentally, estimating quantities like entanglement requires reproducing the same trajectory multiple times, which is exponentially unlikely due to the randomness, a fundamental challenge known as the postselection problem. In realistic implementations, this issue is further complicated by imperfect detector efficiencies, which introduces an intrinsic uncertainty in the trajectory followed. Furthermore, averaging physical quantities over a sample of randomly generated trajectories generally erases all signatures of the phase transition~\cite{ravindranath2023,piroli2023}, posing significant obstacles to the experimental observation of MIPTs.

To address these challenges, here we investigate whether or not signatures of MIPTs persist under \textit{partial} postselection~\cite{minganti2020,kells2023,leung2023,coppola2023,gupta2024,liu2025}, i.e., averaging over a  restricted class of trajectories as opposed to the full ensemble. In other words, we ask if some features of the transition can still be observed when some measurement outcomes are not recorded, resulting in partial loss of information specifying a given trajectory. This approach addresses two critical experimental challenges. First, it allows us to examine the impact of imperfect detection, where the measurement apparatus might fail to detect some events, on the properties of the system. Second, it investigates whether the stringent requirement to focus on a unique trajectory can be relaxed, or if a perfect, resource-expensive postselection is inevitable. A positive answer would potentially offer an experimental advantage in accessing the elusive MIPTs.

In detail, we examine the steady state of a continuously monitored fermionic Kitaev chain, whose evolution is governed by an effective non-Hermitian Hamiltonian with stochastic quantum jumps~\cite{turkeshi2021,paviglianiti2023,piccitto2022}. The postselected trajectory without jumps, known as the no-click limit~\cite{turkeshi2023,biella2021,zerba2023,gopalakrishnan2021,barch2023}, has previously been studied as a simplified model of MIPTs; in this case, the dynamics is non-unitary as a consequence of the measurement backaction, and it drives the system to a pure steady state. We introduce a Liouvillian model that interpolates between this fully postselected scenario and the Lindbladian limit~\cite{minganti2020}, where no postselection occurs and all measurement outcomes are averaged over. By computing the exact fermionic correlators, the entanglement negativity of the steady state, and the Liouvillian gap, we demonstrate that any degree of trajectory averaging destroys the no-click critical phase, thus eliminating the phase transition. As a consequence, the postselection problem cannot be mitigated, at least for the model we consider. We show that imperfect postselection introduces an effective lengthscale that suppresses long-range correlations. Our findings clarify the mechanism by which trajectory averaging obscures measurement-induced phase transitions, and establish a framework for investigating partially postselected monitored dynamics.

The remainder of this manuscript is organized as follows. In Sec.~\ref{s:liouv} we introduce our Liouvillian model of partially postselected monitored dynamics. Section~\ref{s:steady_state_eq} details the derivation of the equation characterizing the steady state and its solution. Then, Sec.~\ref{s:lengthscale} presents the main result of our investigation: we compute the correlation length and demonstrate that only the fully postselected no-click limit can be critical. We substantiate this by computing the entanglement negativity of the steady state and the Liouvillian gap in Sec.~\ref{s:negativity_and_gap}. Finally, we summarize our findings in Sec.~\ref{s:conclusions}

\section{Liouvillian Model}\label{s:liouv}
In this Section we introduce the Liouvillian model we investigate. We consider a fermionic chain of size $L$ undergoing hybrid dynamics that involves unitary Hamiltonian evolution and continuous monitoring. In detail, we consider a weak measurement protocol~\cite{wiseman1996,jacobs2006,svensson2013}: rather than being observed directly, each lattice site is coupled to an ancillary apparatus which is consistently monitored at all times. This process can result in the emission of a fermion from the chain, which is spotted by a detector and will be referred to as a ``click''. Since the losses occur randomly, different realizations of the protocol will give rise to distinct stochastic quantum trajectories, each labelled by the times and positions of its clicks.

We aim to investigate how the properties of the system are affected when the information identifying a specific trajectory is partially lost. To this end, we focus on the so-called no-click trajectory, where each detector reveals no click at any time. We then assume that the monitoring instruments have a limited efficiency, so that they can miss clicks. Specifically, we assume that whenever a click occurs it is recorded only with a certain probability $0\leq q\leq 1$, whereas with probability $1-q$ the event goes unnoticed by the observer (false negative). As a consequence, when detectors report zero clicks the state of the system is not pure, but is instead described by an ensemble of trajectories with undetected events. We show in App.~\ref{a:master_equation} that the evolution of the average density matrix $\hat{\rho}_t$ under this protocol is governed by the master equation
\begin{subequations}\label{liouv_eq}
\begin{equation}\label{master_eq}
    \partial_t \hat{\rho}_t = \mathcal{L}_t \hat{\rho}_t,
\end{equation}
\begin{equation}\label{liouv}
\begin{split}
    \mathcal{L}_t\:\bullet = -i\comm{\hat{H}}{\bullet} + \sum_{j=1}^L\bigg(&(1-q)\hat{L}_j\bullet\hat{L}\daga_j-\frac{1}{2}\acomm{\hat{L}\daga_j\hat{L}_j}{\bullet}\\
    &+q\langle\hat{L}\daga_j\hat{L}_j\rangle_t\bullet\bigg),
\end{split}
\end{equation}
\end{subequations}
where $\hat{H}$ is the Hamiltonian of the model, and $\hat{L}_j = \sqrt{\gamma} \hat{c}_j$ are jump operators describing the on-site fermionic losses. The parameter $\gamma$ is the monitoring rate, quantifying the coupling strength between system and monitored ancillas. In our case, we use the Kitaev chain
\begin{equation}\label{hamiltonian}
    \hat{H} = -\sum_{j=1}^L \left( \hat{c}\daga_j \hat{c}_{j+1} +\hat{c}\daga_j \hat{c}\daga_{j+1} +\mathrm{h.c.} \right) + 2\mu\sum_{j=1}^L \hat{c}\daga_j \hat{c}_j
\end{equation}
with periodic boundary conditions. Notice that Eq.~\eqref{liouv_eq} is non-linear, as it involves the expectation value $\langle\hat{L}\daga_j\hat{L}_j\rangle_t = \Tr(\hat{\rho}_t\hat{L}\daga_j\hat{L}_j)$. This term guarantees a trace-preserving evolution, and arises naturally from the derivation. The Liouvillian of Eq.~\eqref{liouv} is called quasi-free because the Hamiltonian is quadratic and the jump operators are linear in the fermionic operators. Notably, it can be solved exactly by diagonalizing it~\footnote{To be precise, its eigenoperators can be computed exactly, whereas its eigenvalues are determined up to the constant term $q\langle\hat{L}_j\daga\hat{L}_j\rangle_t$ that depends on the current state.} through the formalism of third quantization~\cite{prosen2008,prosen2010,barthel2022}. Most importantly, it can be proven that the dynamics generated by quasi-free Liouvillians preserves the Gaussianity of the state~\cite{coppola2023}, and thus the latter is fully characterized by its correlation matrix~\cite{surace2022}.

There are two notable limiting cases of the Liouvillian. For $q=0$, Eq.~\eqref{liouv_eq} reduces to a regular Lindblad master equation, which describes the evolution obtained by averaging accross all possible, randomly generated, trajectories. In this case, it can be proven that the steady state can never be critical~\cite{zhang2022}, and must have a finite correlation length. In the opposite limit of $q=1$,  each trajectory with one or more clicks is discarded, leaving only the no-click trajectory. This case is described by a deterministic pure state evolution, governed by a non-Hermitian Schrödinger equation~\cite{wiseman1996,jacobs2006}. The no-click limit of this model has been studied extensively in the literature, and at long times it is known to manifest an extended critical phase with long-range correlations for $|\mu|<1$ and $\gamma<\gamma_c(\mu) = 4\sqrt{1-\mu^2}$~\cite{turkeshi2021,biella2021,turkeshi2023, DiFresco2024}. By tuning the parameter $q$ we can interpolate between these two extreme cases. The main goal of our investigation is to understand whether the critical properties of the no-click limit are retained for $q<1$. As detailed in the subsequent sections, we anticipate a negative answer: for all $q<1$, we observe a finite correlation length and the absence of criticality.



The paradigm of partial postselection has already been considered by Refs.~\cite{kells2023,leung2023}. These works focus on quantities that are non-linear in the density matrix, and compute their averages over a restricted ensemble of trajectories. In our case, we are instead interested in linear observables and in the properties of the mixed steady state.

An alternative but equivalent interpretation that leads to Eq.~\eqref{liouv_eq} is the following. We assume that quantum trajectories are sampled regularly without any postselection, and we bias this process by discarding some realizations. For a randomly generated trajectory, we introduce a probability $q$ to drop it whenever a click occurs, meaning it is completely disregarded and excluded from the ensemble over which observables are computed. This biased sampling protocol yields the same Liouvillian (see App.~\ref{a:master_equation}) of the partial postselection paradigm, and constitutes an experimentally viable way to realize it. In both interpretations, the parameter $q$ penalizes the relative probability of trajectories with many clicks, making them less likely to occur.

\section{Steady-State Correlation Matrix}\label{s:steady_state_eq}
We now move to the main focus of our study, which regards the investigation of the steady-state properties of Eq.~\eqref{liouv_eq}. By exploiting the property that the Liouvillian $\mathcal{L}_t$ of Eq.~\eqref{liouv} preserves the Gaussianity of the state, we derive an algebraic Riccati equation (ARE) that yields the exact correlation matrix of the steady state. We point out that Liouvillians similar to that considered here have been recently studied in Ref.~\cite{coppola2023} by using related techniques. The authors of that work characterize dynamical aspects of the model by computing quantities like the click waiting-time distribution and the no-click probability. In contrast, we work directly with the steady state to study its spatial correlations and entanglement properties.

From here on, it is convenient to work using the Majorana fermionic operators defined by $\hat{w}_{j,1} = (\hat{c}_j+\hat{c}\daga_j)/\sqrt{2}$, $\hat{w}_{j,2} = -i (\hat{c}_j-\hat{c}\daga_j)/\sqrt{2}$. The Hamiltonian and jump operators are rewritten as
\begin{equation}\label{hamiltonian_majorana}
    \hat{H} = \sum_{m,n=1}^L\sum_{\mu,\nu=1}^2 \mathbb{H}_{(m,\mu),(n,\nu)}\hat{w}_{m,\mu}\hat{w}_{n,\nu},
\end{equation}
where $\mathbb{H} = -\mathbb{H}^T$~\footnote{There are multiple equivalent ways to define $\mathbb{H}$. The requirement of $\mathbb{H} = -\mathbb{H}^T$ selects a unique one, and is convenient for later calculations.}, and 
\begin{equation}
    \hat{L}_j = \sum_{m=1}^L\sum_{\mu=1}^2\ell^{(j)}_{m,\mu}\hat{w}_{m,\mu}.
\end{equation}
For later convenience, let us introduce the so-called bath matrix
\begin{equation}\label{bath_matrix}
    M_{(m,\mu),(n,\nu)} = \sum_{j=1}^L \ell^{(j)}_{m,\mu}\left(\ell^{(j)}_{n,\nu}\right)^*,
\end{equation}
which is easily shown to be Hermitian. We are interested in computing the correlation matrix
\begin{equation}\label{gamma}
    \Gamma_{(m,\mu),(n,\nu)} = \frac{i}{2}\Tr\left(\hat{\rho}\comm{\hat{w}_{m,\mu}}{\hat{w}_{n,\nu}}\right)
\end{equation}
in the steady state of the dynamics. To achieve this, we first use Eq.~\eqref{liouv_eq} to obtain an equation of motion for $\Gamma$. In the process one finds that the equation involves expectation values of products of $4$ Majorana operators. By exploiting the Gaussianity of the density matrix $\hat{\rho}_t$ at all times, we can apply Wick's theorem to reduce these fourth order terms, and obtain a closed equation for $\Gamma$. After a tedious but straightforward calculation, this leads to 
\begin{equation}\label{time_ARE}
    \partial_t\Gamma_t = X \Gamma_t + \Gamma_t X^T + Y + \Gamma_t Z \Gamma_t,
\end{equation}
where the matrices $X$, $Y$, and $Z$ are given by
\begin{subequations}
    \begin{eqnarray}
        X &=& -2i\mathbb{H} - (1-q)\Re M,\\
        Y &=& (1-q/2)\Im M,\\
        Z &=& 2q\Im M.
    \end{eqnarray}
\end{subequations}

We can exploit translational symmetry to simplify the problem by moving to momentum space. For $A\in\left\{\Gamma,X,Y,Z\right\}$, the matrix elements $A_{(m,\mu),(n,\nu)} = A_{\mu,\nu}(m-n)$ are functions of the distance $m-n$. Introducing the momentum-space matrices $\Tilde{A}_{\mu,\nu}(k) = \sum_{x=1}^L e^{-i k x}A_{\mu,\nu}(x)$, Eq.~\eqref{time_ARE} yields $L$ independent equations for each $k$-mode. Finally, by imposing $\partial_t\Tilde{\Gamma}_t(k) = 0$ we obtain the steady-state equation
\begin{equation}\label{ARE}
    \Tilde{X}(k)\Tilde{\Gamma}(k) \!+\! \Tilde{\Gamma}(k) \Tilde{X}^T(-k) \!+\! \Tilde{Y}(k) \!+\! \Tilde{\Gamma}(k)\Tilde{Z}(k)\Tilde{\Gamma}(k) = 0,
\end{equation}
which has the form of a $2\times 2$ ARE.

Here, we briefly outline the procedure to solve the ARE, deferring a more detailed discussion to App.~\ref{a:solve_ARE}. In general, Eq.~\eqref{ARE} can be solved by diagonalizing a $4\times 4$ block matrix, and it has six distinct solutions for each $k$. Most of these do not correspond to physically meaningful correlation matrices, as they do not fulfill the properties $\Tilde{\Gamma}_{1,1}(k) = -\Tilde{\Gamma}_{2,2}(k)$ and $\Tilde{\Gamma}(k) = -\Tilde{\Gamma}\daga(k)$. The former is due to a symmetry of the Liouvillian, while the latter follows from the skew-symmetry of fermionic correlation matrices [see Eq.~\eqref{gamma}]. The above constraints exclude four unphysical solutions. In addition, they bound $\Tilde{\Gamma}(k)$ to be parameterized by only three real variables, which we then compute analytically. Of the remaining two solutions we then single out the one that yields the correct limits for $q\to 0$ and $q\to 1$.

\begin{figure}[ht]
\centering
 \includegraphics[width=\columnwidth]{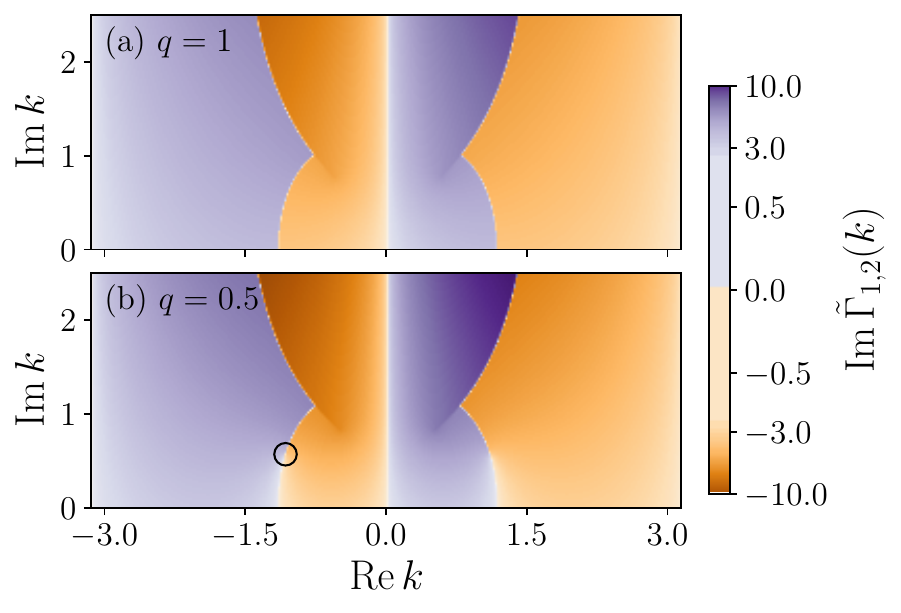}
 \caption{Profile of the correlation function $\Im \Tilde{\Gamma}_{1,2}$ for complex argument $k$, using $\mu = 0.4$, $\gamma=1$, and (a) $q=1$, (b) $q=0.5$. In the no-click limit (a) the discontinuous line extends to $\Im k = 0$, whereas in (b) it terminates at the point highlighted by the black circle.}
 \label{f:landscape}
\end{figure}

\section{Lengthscale of correlations}\label{s:lengthscale}
We now leverage the previous formalism to study the critical properties of our model. Using complex analysis tools, we are able to set an upper bound to the lengthscale $\xi$ of the correlation functions, valid in the thermodynamic limit. This allows us to demonstrate that $\xi$ is always finite for $q<1$, which implies that any amount of partial averaging over trajectories forbids algebraically decaying correlations with the distance.
\begin{figure}[t]
\centering
 \includegraphics[width=\columnwidth]{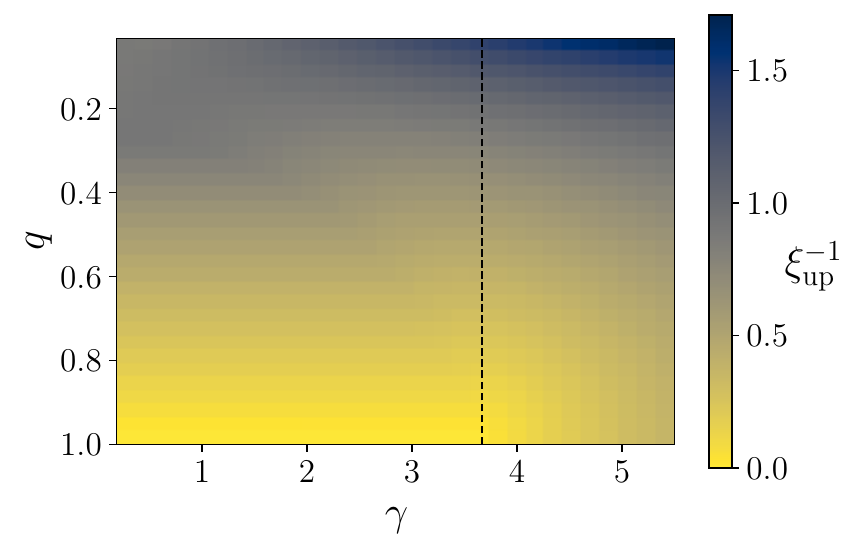}
 \caption{Inverse upper correlation length $\xi_\text{up}^{-1}$ for $\mu=0.4$, as a function of $\gamma$ and $q$. The dashed line indicates the critical $\gamma_c(\mu)$ of the no-click limit. We observe $\xi_\text{up}^{-1}>0$ for any $q<1$.}
 \label{f:xi_up}
\end{figure}
\indent A useful approach to establish whether a correlation function decays as an exponential or as a power law consists in studying the singularities of its Fourier transform in the complex plane. We promote the momentum $k$ to a complex variable $k\in\mathbb{C}$, and we look for non-analiticities of $\Tilde{\Gamma}(k)$. Two examples are provided in Fig.~\ref{f:landscape}, where $\Im \Tilde{\Gamma}_{1,2}(k)$ is displayed (a) in the no-click limit ($q=1$) for critical values of the parameters, and (b) for $q<1$, with all the other parameters left unchanged. The correlation matrix features lines of discontinuity rather than isolated poles. We denote by $\mathcal{C}$ the collection of these lines for $\Im k \geq 0$. For $q=1$ the lines touch the real axis $\Im k=0$, which is a necessary condition for a diverging correlation length, as shown later. In contrast, for $q<1$ we observe that $\Tilde{\Gamma}(k)$ is regular on the real axis. In this case, we can write the real-space correlation matrix $\Gamma(x)$ as a contour integral. After performing the change of variable $z=e^{ik}$, we obtain
\begin{equation}\label{contour_int}
    \Gamma(x) = \frac{1}{2\pi i} \oint_{\mathcal{S}}z^{x-1} \Tilde{\Gamma}(k(z))dz ,
\end{equation}
where $\mathcal{S}$ is the unit circle centered at the origin, encircling the singular lines $\mathcal{C}$ mapped to the new variable $z$. Proceeding as detailed in App.~\ref{a:complex_analysis}, it is straightforward to prove that
\begin{equation}\label{corr_bound}
    |\Gamma(x)|\leq C \exp \left(-\min_{k\in\mathcal{C}}\left\{\Im k\right\} x\right),
\end{equation}
where $C$ is a constant and $\min_{k\in\mathcal{C}}\left\{\Im k\right\}$ is the imaginary part of the closest point of $\mathcal{C}$ to the real axis. This finally sets the upper bound
\begin{equation}\label{xi_up}
    \xi \leq \xi_\text{up} = \left(\min_{k\in\mathcal{C}}\left\{\Im k\right\}\right)^{-1}.
\end{equation}

We can now study the correlation length of the Liouvillian steady state by looking for the singular lines $\mathcal{C}$ and computing their minimal distance from the real axis. Fig.~\ref{f:xi_up} shows the upper lengthscale $\xi_\text{up}$ as a function of $q$ and $\gamma$, fixing the chemical potential $\mu$. In general, we observe that for all parameter choices the lengthscale is a monotonically increasing function of $q$. This indicates that the less postselection is performed, the less information is retained on the long-range behavior of correlations. In particular, we observe that $\xi_{up}$ can diverge only for $q=1$, which demonstrates that the steady states for $q<1$ are not critical. Specifically, for $\gamma<\gamma_c(h)$ we can evaluate the correlation length explicitly for $1-q\ll 1$, obtaining $\xi_\text{up}^{-1}=1-q + \mathcal{O}((1-q)^2)$. This result can be proven exactly by inspecting the analytic expression of $\Tilde{\Gamma}_k$, as shown in App.~\ref{a:solve_ARE}.

\begin{figure}[ht!]
\centering
 \includegraphics[width=0.76\columnwidth]{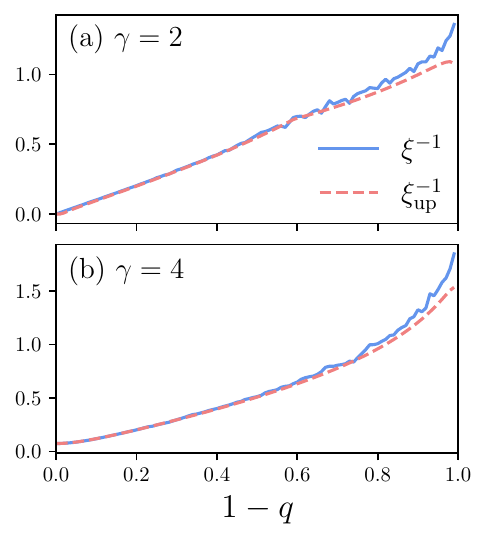}
\caption{Decay of the correlation length induced by partial averaging, as a function of the detector inefficiency $1-q$. We set $\mu=0.4$ and  (a) $\gamma=2<\gamma_c(\mu)$, (b) $\gamma=4>\gamma_c(\mu)$, corresponding to critical and off-critical no-click limits, respectively. $\xi$ is fitted from the correlation matrix computed for a system size of $L=2048$, whereas $\xi_\text{up}$ is the theoretical upper bound of Eq.~\eqref{xi_up}.}
 \label{f:inverse_xi}
\end{figure}
Finally, we also evaluated the true correlation length $\xi$ by reconstructing $\Gamma(x)$ and fitting its decay~\footnote{We assume the form $|\Gamma_{\mu,\nu}| \approx A e^{-|x|/\xi}/|x|^\alpha$ and optimize the three parameters $A$, $\alpha$, and $\xi$. We point out that the exact correlation functions manifest oscillations as well, but these are treated as noise around the previous decay profile.}. Figure~\ref{f:inverse_xi} shows a comparison between $\xi_\text{up}$ and the fitted $\xi$. Despite the former being only an upper bound, we observe that it approximates quite well the true correlation length. At small values of $q$ the estimation becomes noisy because we
are fitting an exponential decay with very small length-
scale $\xi\sim 1$.

\section{Entanglement and spectral gap}\label{s:negativity_and_gap}
We complement the previous findings by computing additional quantities that are indicators of critical behavior. First, we study the entanglement negativity of the steady state and show that it provides similar information to the above analysis. We then consider the Liouvillian gap of the spectrum of Eq.~\eqref{liouv}, and observe that the model can becomes gapless only in the no-click limit of $q\to 1$.
\begin{figure}[t]
\centering
 \includegraphics[width=0.85\columnwidth]{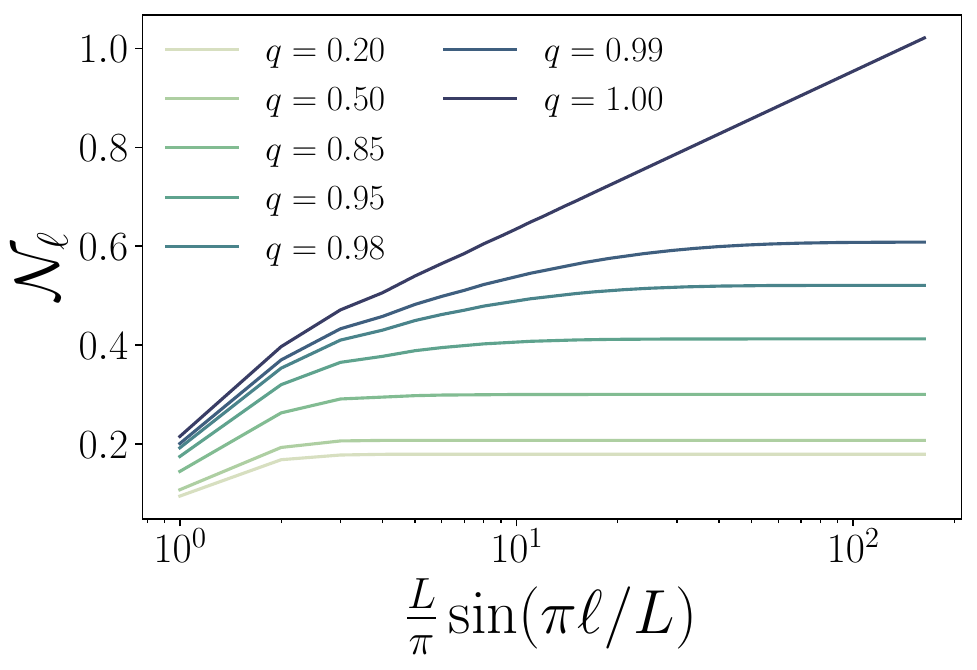}
 \caption{Scaling of the entanglement negativity $\mathcal{N}_\ell$ as a function of the chord length $\frac{L}{\pi}\sin(\pi\ell/L)$~\cite{calabrese2004}. Here we use $L=512$, $\mu=0.4$, and $\gamma=1$.}
 \label{f:negativity}
\end{figure}

The entanglement negativity is a measure of bipartite entanglement that, as opposed to the entanglement entropy, has the advantage of detecting genuine quantum correlations also for mixed states~\cite{vidal2002,plenio2005,amico2008,horodecki2009,calabrese2012}. Given a density matrix $\hat{\rho}_{A,B}$ describing two non-overlapping subsystems $A$ and $B$, the negativity of a fermionic system~\cite{shapourian2017,shapourian2019,murciano2021,murciano2022,rottoli2023} is formally given by
\begin{equation}
    \mathcal{N}_{A,B} = \log \Tr \left|\hat{\rho}_{A,B}^{\Tilde{T}_A}\right|,
\end{equation}
where $\hat{\rho}_{A,B}^{\Tilde{T}_A}$ denotes the twisted partial transpose of $\hat{\rho}_{A,B}$ with respect to subsystem $A$. For Gaussian states, this quantity can be computed efficiently using the methods described in Refs.~\cite{shapourian2019,murciano2021}. In our case, we pick $A$ to be a compact subsystem of length $\ell$ and $B$ to be the rest of the chain. We show our results in Fig.~\ref{f:negativity}. For $q=1$ we observe an unbounded logarithmic growth, which is expected due to the critical nature of the state for the chosen parameters. In contrast, for any $q<1$ $\mathcal{N}_\ell$ eventually saturates to a constant. This is expected for states with a finite correlation length: the ground states of gapped one-dimensional models manifest an initial logarithmic growth $\sim \log \ell$ for $\ell \lesssim \xi$, and an eventual saturation to $\sim \log \xi$ at larger subsystem sizes~\cite{calabrese2013}. The entanglement negativity shows that the steady state has quantum correlations, but these get smaller and smaller as more trajectories are considered in the averaging.

\begin{figure}[t!]
\centering
 \includegraphics[width=\columnwidth]{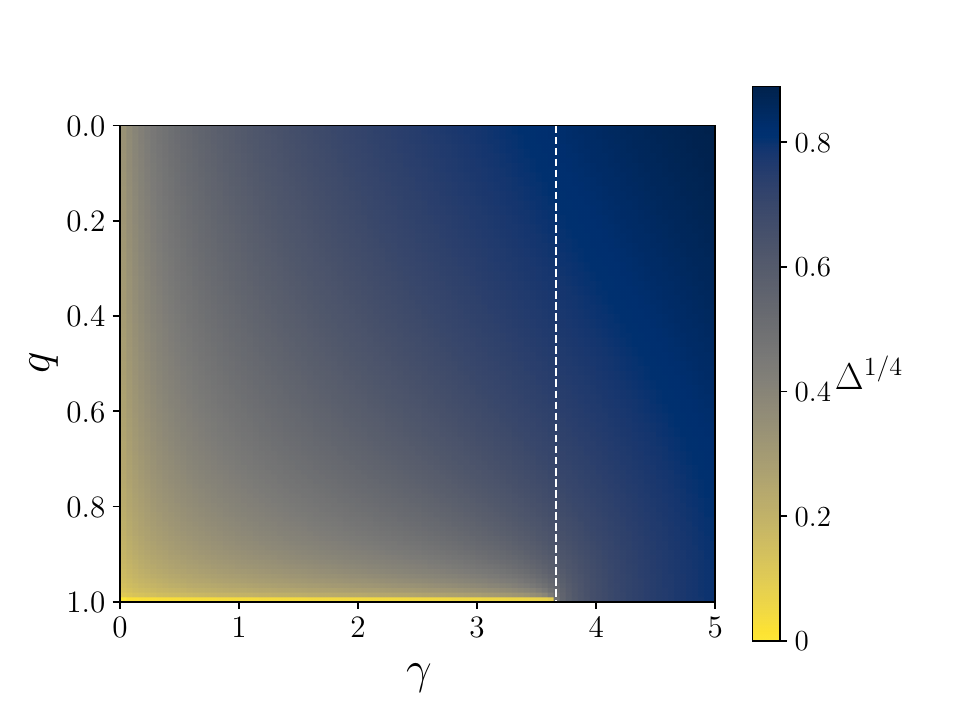}
 \caption{Liouvillian gap as a function of $q$ and $\gamma$ for $L=128$ and $\mu=0.4$. We display $\Delta^{1/4}$ instead of $\Delta$ to improve graphical visibility. The dashed line indicates the critical $\gamma_c(\mu)$ of the no-click limit. For any $q<1$ the gap is always finite.}
 \label{f:gap}
\end{figure}

Phase transitions in open quantum systems can also be characterized through the Liouvillian spectral gap~\cite{kessler2012,horstmann2013,minganti2018}. Denoting by $\lambda_i$, $i=0,\dots,4^N-1$, the Liouvillian eigenvalues and ordering them such that $\Re \lambda_0\geq \Re \lambda_1 \geq \dots \geq \Re \lambda_{4^N-1}$, the Liouvillian gap is defined as $\Delta = \Re (\lambda_1-\lambda_0)$. Since $\lambda_0$ corresponds to the slowest-decaying mode, the gap quantifies how quickly it is reached. The closure of the Liouvillian gap is a signature of a phase transition. We can evaluate it numerically by diagonalizing the Liouvillian using the formalism of third quantization, which is discussed in App.~\ref{a:third_quantization}. In Fig.~\ref{f:gap} we show $\Delta$ as a function of the parameters of our model. The gap closes only in the no-click limit of $q=1$, thus confirming that no phase transition can occur for imperfect postselection corresponding to $q<1$.

\section{Conclusions}\label{s:conclusions}
This work explores the impact of imperfect detection and partial postselection on quantum correlations in an exactly-solvable model of monitored dynamics. By computing the exact correlation matrix and employing complex analysis methods, we establish an upper bound to the correlation length of the system, which closely approximates the exact one. We reveal that trajectory averaging introduces an effective lengthscale for correlation functions, causing the entanglement phase transition observed in the fully postselected no-click limit to vanish for any degree of information loss ($q<1$). This conclusion is further supported by our analysis of the entanglement negativity and the Liouvillian gap. Although non-trivial correlations persist at lengthscales below $\xi$, the latter diminishes rapidly as $q$ decreases. 

A crucial question following our findings is whether this result generalizes beyond our model. Our Liouvillian is integrable, and it may not be representative of generic monitored models. We expect that the appearance of a $q$-dependent correlation length holds quite generally, in line with the common understanding that decoherence destroys long-range quantum correlations. Still, there might exist models featuring an extended critical region with $\xi^{-1}=0$ above some $q_c<1$. Future investigations could study the role of partial postselection in systems with Hamiltonian interactions or more complex jump operators, possibly producing non-Gaussian dynamics. Beyond Gaussianity, the problem becomes much harder to approach, as analytical tools are limited. At the same time, numerical simulations based on quantum trajectory unravelings are always restricted to finite system sizes, making it challenging to distinguishing between a very large $\xi$ and a formally diverging one. In this context, the framework of the algebraic Riccati equation for the steady-state correlation matrix could be valuable for developing a perturbative approach in the presence of weak integrability-breaking terms.

Another important consideration is to what extent the instability of the no-click phase transition depends on the criterion used to discard quantum trajectories: in our study, partial postselecion excludes trajectories with many jumps, but alternative criteria could be explored.

Finally, it would be interesting to investigate the role of dimensionality. Unlike fermions, monitored bosonic systems in two or more dimensions can exhibit criticality even without postselection~\cite{zhang2022}. In such cases, one might expect a non-trivial phase diagram at $0<q<1$, bridging the Lindbladian and no-click limits. These intriguing questions are left for future works.

\textbf{Acknowledgements} --- A.S. would like to acknowledge support from PNRR MUR project PE0000023-NQSTI and Quantera project SuperLink. A.C. acknowledges support from European Union – Next Generation EU through Project Eurostart 2022 “Topological atom-photon interactions for quantum technologies" (MUR D.M. 737/2021) and through Project PRIN 2022-PNRR no. P202253RLY “Harnessing topological phases for quantum technologies". A.C. and D.V. acknowledge support from European Union - Next Generation EU through project THENCE - Partenariato Esteso NQSTI - PE00000023 - Spoke 2.
 
\appendix

\section{Master Equation with Partial Postselection}\label{a:master_equation}
In this Appendix we derive the master equation of Eq.~\eqref{liouv_eq} for the partially postselected monitored dynamics protocol. Before providing the derivation, we first summarize the framework of positive-operator valued measures (POVMs), which can be used to describe the weak measurements we consider. In fact, as mentioned in the main text, we assume that the system is not measured directly, but rather it is coupled to ancillary degrees of freedom that are monitored instead. This corresponds to a generalized measurement, which can be characterized using the theory of POVMs.

The action of a generalized measurement with $M$ possible outcomes is described by $M$ Kraus operators $\hat{A}_m$, $m=1,\dots,M$, satisfying the identity
\begin{equation}\label{kraus_sum}
    \sum_m \hat{A}\daga_m\hat{A}_m = \mathds{1}.
\end{equation}
Each $\hat{A}_m$ characterizes how the measurement acts on the state when outcome $m$ is observed. In detail, given a pre-measurement state $\ket{\psi}$, the post-measurement state $\ket{\psi_m}$ corresponding to outcome $m$ is given by
\begin{equation}
    \ket{\psi}_m = \frac{\hat{A}_m\ket{\psi}}{\sqrt{\bra{\psi}\hat{A}\daga_m\hat{A}_m\ket{\psi}}}.
\end{equation}
The observed outcome is random, and the probability of obtaining $m$ is given by $p_m=\bra{\psi}\hat{A}\daga_m\hat{A}_m\ket{\psi}$.

In our case, on each site $j$ we consider a measurement with two possible outcomes $0$ and $1$, corresponding to having no clicks or a click, respectively. Since a click corresponds to the loss of a fermion from the chain, we use $\hat{A}_{j,1} = \sqrt{p}\hat{c}_j$, where the parameter $p$ quantifies the probability of this outcome. Specifically, since we want to model a continuous monitoring process where measurements are performed at each infinitesimal timestep $dt$, we pick $p = \gamma dt$ in such a way that the average number of clicks scales linearly in time. This yields $\hat{A}_{j,1} = \sqrt{\gamma dt} \hat{c}_j \equiv \sqrt{dt}\hat{L}_j$, where $\hat{L}_j$ is the jump operator that appears in the master equation of Eq.~\eqref{liouv_eq}. There are infinitely many choices for the other Kraus operator $\hat{A}_{j,0}$, and we use
\begin{equation}\label{kraus_0}
    \hat{A}_{j,0} = e^{-\frac{dt}{2} \hat{L}\daga_j\hat{L}_j},
\end{equation}
which fulfills the constraint of Eq.~\eqref{kraus_sum} at $\mathcal{O}(dt)$.

We can now derive Eq.~\eqref{liouv_eq} explicitly. For simplicity, assume that a single site $j$ is monitored. Let $\hat{\rho}_t$ be the average density matrix at time $t$. With probability $q$, the measurement is postselected to have the no-click outcome, meaning that the state will evolve to $\hat{A}_{j,0}\hat{\rho}_t\hat{A}\daga_{j,0}/p_0$. Notice that $p_0=\mathcal{O}(1)$, ensuring that this state is well-defined. In contrast, with probability $1-q$ no postselection is performed and both outcomes are possible; in this case, the evolved state will be given by $\hat{A}_{j,m}\hat{\rho}_t\hat{A}\daga_{j,m}/p_m$, where $m=0,1$ is picked randomly according to the Born rule probabilities $p_m$. Overall, the average density matrix at time $t+dt$ is a mixture of all possible final states, each weighted with the corresponding probability of occurring, thus yielding
\begin{equation}\label{mixture}
\begin{split}
    \hat{\rho}_{t+dt} = &q \frac{\hat{A}_{j,0}\hat{\rho}_t\hat{A}\daga_{j,0}}{p_0}\\
    &+ (1-q)\left[p_0\frac{\hat{A}_{j,0}\hat{\rho}_t\hat{A}\daga_{j,0}}{p_0} + p_1\frac{\hat{A}_{j,1}\hat{\rho}_t\hat{A}\daga_{j,1}}{p_1}\right].
\end{split}
\end{equation}
By expanding $\hat{A}_{j,0}\approx \hat{\mathds{1}}-\frac{dt}{2}\hat{L}\daga_j \hat{L}_j$ in Eq.~\eqref{mixture} and keeping only the leading order in $dt$, we finally get
\begin{equation}
\begin{split}
    \hat{\rho}_{t+dt} = &\hat{\rho}_t -\frac{dt}{2}\acomm{\hat{L}\daga_j\hat{L}_j}{\hat{\rho}_t}+q dt \langle\hat{L}\daga_j\hat{L}_j\rangle_t\hat{\rho}_t\\
    &+(1-q) dt \hat{L}_j\hat{\rho}_t\hat{L}_j,
\end{split}
\end{equation}
and by reordering terms we arrive at
\begin{equation}\label{single_site_liouv}
    \partial_t\hat{\rho}_t = -\frac{1}{2}\acomm{\hat{L}\daga_j\hat{L}_j}{\hat{\rho}_t}+q \langle\hat{L}\daga_j\hat{L}_j\rangle_t\hat{\rho}_t+(1-q) \hat{L}_j\hat{\rho}_t\hat{L}_j.
\end{equation}
Notice that the contribution $q \langle\hat{L}\daga_j\hat{L}_j\rangle_t\hat{\rho}_t$, which ensures trace preservation, naturally appears from the derivation from the expansion of $1/p_0$ at leading order. We point out that the same equation is obtained by assuming that the trajectory with $m=1$ is discarded with a probability $q$. In this case one has a probability $p_0$ to observe $m=0$ and a probability $(1-q)p_1$ to obtain $m=1$. The density matrix must then be renormalized so that $\Tr\left(\hat{\rho}_{t+dt}\right)=1$, as discarding trajectories results in a loss of probability.

For measurements acting on all sites, since they act independently, we simply need to add a sum over the lattice index $j$ on the right-hand side of Eq.~\eqref{single_site_liouv}. Finally, by including the coherent Hamiltonian evolution term we obtain Eq.~\eqref{liouv_eq}.

\section{Solution of the ARE}\label{a:solve_ARE}
We now describe the general method to solve the ARE of Eq.~\eqref{ARE}, and then provide the exact solution for our problem. First, we introduce the block matrix
\begin{equation}
    Q = \begin{pmatrix}
        \Tilde{X}^T(-k) & \Tilde{Z}(k)\\
        -\Tilde{Y}(k) & -\Tilde{X}(k)
    \end{pmatrix},
\end{equation}
which allows us to rewrite the ARE in matrix form as
\begin{equation}\label{matrix_ARE}
    \begin{pmatrix}
        \Tilde{\Gamma}(k) & -\mathds{1}
    \end{pmatrix} Q \begin{pmatrix}
        \mathds{1} \\
        \Tilde{\Gamma}(k)
    \end{pmatrix} = 0.
\end{equation}
Let $W$ be the solution of the eigenvalue equation $Q$, i.e., $Q W = W \Lambda$, where $\Lambda = \text{diag}\{\lambda_1,\lambda_2,\lambda_3,\lambda_4\}$ contains the egenvalues of $Q$. The columns of $W$ are the right eigenvectors of $Q$. If we rewrite $W$ as a block matrix
\begin{equation}
    W = \begin{pmatrix}
        W_{1,1} & W_{1,2} \\
        W_{2,1} & W_{2,2}
    \end{pmatrix},
\end{equation}
where each block $W_{m,n}$ has size $2\times 2$, the solution of the ARE is given by
\begin{equation}\label{matrix_ARE_solution}
    \Tilde{\Gamma}(k) = W_{2,1}W_{1,1}^{-1}.
\end{equation}
This can be checked by inserting this ansatz in Eq.~\eqref{matrix_ARE} and using the identity $Q = W \Lambda W^{-1}$. Equation~\eqref{matrix_ARE_solution} involves only two columns of $W$, i.e., only two eigenvectors of $Q$. Any reordering of the eigenvectors provides a valid solution of the ARE. Since the number of inequivalent permutations is six, we conclude that there are these many solutions.

As mentioned in the main text, the physically meaningful solution must satisfy two properties, which we now prove. Let us start by showing that $\Tilde{\Gamma}_{1,1}(k) = -\Tilde{\Gamma}_{2,2}(k)$. This is a consequence of a symmetry of the Liouvillian under a swap of type-$1$ and type-$2$ Majorana operators plus an inversion of the lattice chain. Denoting by $\mathcal{L}[\{\hat{w}_{j,1}\},\{\hat{w}_{j,2}\}]$ the Liouvillian written in terms of the Majorana operators, it is easily checked that it satisfies $\mathcal{L}[\{\hat{w}_{j,1}\},\{\hat{w}_{j,2}\}] = \mathcal{L}[\{\hat{w}_{L-j,2}\},\{-\hat{w}_{L-j,1}\}]$. It follows that $\Gamma_{1,1}(x)=\Gamma_{2,2}(-x)$. Combining this with the identity $\Gamma_{\mu,\nu}(x) = -\Gamma_{\nu,\mu}(-x)$, which is a direct consequence of its definition, we obtain $\Gamma_{1,1}(x)=-\Gamma_{2,2}(x)$, and thus $\Tilde{\Gamma}_{1,1}(k) = -\Tilde{\Gamma}_{2,2}(k)$. Moving on, let us now prove that $\Tilde{\Gamma}(k)$ must be an anti-Hermitian matrix. This is easily seen by rewriting
\begin{equation}
\begin{split}
    (\Tilde{\Gamma}\daga(k))_{\mu,\nu} &= \Tilde{\Gamma}^*_{\nu,\mu}(k) = \sum_m e^{i k x} \Gamma^*_{\nu,\mu}(m)\\
    &= -\sum_m e^{i k x} \Gamma_{\mu,\nu}(-m) = -\Tilde{\Gamma}_{\mu,\nu}(k),
\end{split}
\end{equation}
where we used $\Gamma = -\Gamma\daga$, following directly from its definition.

The most general traceless anti-Hermitian matrix can be parameterized as
\begin{equation}
    \Tilde{\Gamma}(k) = i A \begin{pmatrix}
        1 & a+i b\\
        a-i b & -1
    \end{pmatrix},
\end{equation}
where $A,a,b\in\mathbb{R}$. By inserting this ansatz in the ARE we obtain three independent equations to determine these parameters, allowing to obtain their analytic expressions. This approach produces two solutions $\Tilde{\Gamma}_\pm$ to the problem, corresponding to the parameters $\{A_\pm,a_\pm,b_\pm\}$. For our Liouvillian, we explicitly have
\begin{subequations}
\begin{equation}
    \Tilde{X}(k) = \begin{pmatrix}
        -(1-q)\frac{\gamma}{2} & 2\mu - 2 e^{-i k} \\
        -2\mu + 2 e^{i k} & -(1-q)\frac{\gamma}{2}
    \end{pmatrix},
\end{equation}
\begin{equation}
    \Tilde{Y}(k) = -(1-\frac{q}{2})\frac{\gamma}{2}\begin{pmatrix}
        0 & 1 \\
        -1 & 0
    \end{pmatrix},
\end{equation}
\begin{equation}
    \Tilde{Z}(k) = -q \gamma\begin{pmatrix}
        0 & 1 \\
        -1 & 0
    \end{pmatrix}.
\end{equation}
\end{subequations}
Defining for convenience
\begin{subequations}
    \begin{equation}
        R = 2\mu - 2 \cos k,
    \end{equation}
    \begin{equation}
        I = 2 \sin k,
    \end{equation}
    \begin{equation}
    S = \sqrt{\gamma^4 + (R^2+I^2)^2 + 8\gamma^2[(1-4q+2q^2)I^2+R^2]},
\end{equation}
\end{subequations}
we find
\begin{subequations}
\begin{equation}\label{a_pm}
    a_\pm = \pm \frac{2\sqrt{2}R}{\sqrt{\gamma^2-4(R^2+I^2)+S}},
\end{equation}
\begin{equation}\label{b_pm}
    b_\pm = -\frac{1+a_\pm^2}{a_\pm}\frac{R}{I},
\end{equation}
\begin{equation}\label{A_pm}
    A_\pm = \frac{(1-q)\gamma a_\pm + 2R }{2q\gamma(1+a_\pm^2)R}I.
\end{equation}
\end{subequations}

At last, we need to pick the correct solution between the two. To do so it is useful to consider the limits of $q\to 0$ and $q\to 1$. In the latter, the analytic expression of the correlation matrix is already available~\cite{turkeshi2023,paviglianiti2023,zerba2023} and it coincides with $\Tilde{\Gamma}_-$. We conclude that $\Tilde{\Gamma}_-$ is the physically meaningful solution of the ARE describing the steady state. This is further established by looking at the limit of $q\to 0$, in which $A_-$ behaves regularly while $A_+$ is singular, diverging as $1/q$.

Having the exact expression of $\Tilde{\Gamma}(k)$, we can prove that it cannot manifest singular behavior for $k\in\mathbb{R}$ if $q<1$. To do so, it is convenient to analyze the no-click limit of $q=1$. In this case, for $|\mu|<1$ the momentum $k^* = \arccos \mu$ is such that $R = 0$. It is easy to check that $A_-$, $a_-$, and $b_-$ are regular functions of $k$ at $k^*$ if $\gamma>\gamma_c(\mu)=4\sqrt{1-\mu^2}$. In contrast, for $\gamma<\gamma_c(\mu)$ one can check that $a_-\sim \mathrm{Sign}(k-k^*)$, $b_-\sim |k-k^*|$, whereas $A_-$ is still regular. These behaviors are responsible for the algebraic correlations observed in the no-click limit. In the case of $q<1$, $a_-$ can be singular only if its denominator vanishes at $k^*$. This yields the condition
\begin{equation}
    \left[S^2- \left(\gamma^2-4(R^2+I^2)\right)^2\right]\bigg|_{k^*} = 0,
\end{equation}
which is satisfied only at $q=1$. We thus conclude that $a_-$ is regular at $k^*$ for $q<1$, and specifically we have $a_- \sim R$. From Eqs.~\eqref{b_pm} and~\eqref{A_pm} we then see that also $b_-$ and $A_-$ must be regular at $k^*$.

Beyond observing that $\xi_\text{up}^{-1}>0$ for $q<1$, we can actually evaluate it explicitly for $1-q\ll 1$. In this limit and for $\gamma<\gamma_c(h)$, the non-analitic (complex) $k$-point with smallest imaginary part will stay close to $k^*$. By assuming $k = k^* + dk$, where $dk=\mathcal{O}(1-q)$, and evaluating the denominator of Eq.~\eqref{a_pm} at leading order in $1-q$, we obtain
\begin{equation}
 a_- \propto \frac{1}{\sqrt{(1-q)^2+(dk)^2}}. 
\end{equation}
This function develops a jump discontinuity for $|dk|>1-q$, caused by the branch cut of the square root. Since this condition can be met only for $\Im dk >1-q$, we conclude that $\xi_\text{up}^{-1} = 1-q + \mathcal{O}((1-q)^2)$.

\section{Complex space analysis of the correlation length}\label{a:complex_analysis}
\begin{figure}[ht]
\centering
 \includegraphics[width=\columnwidth]{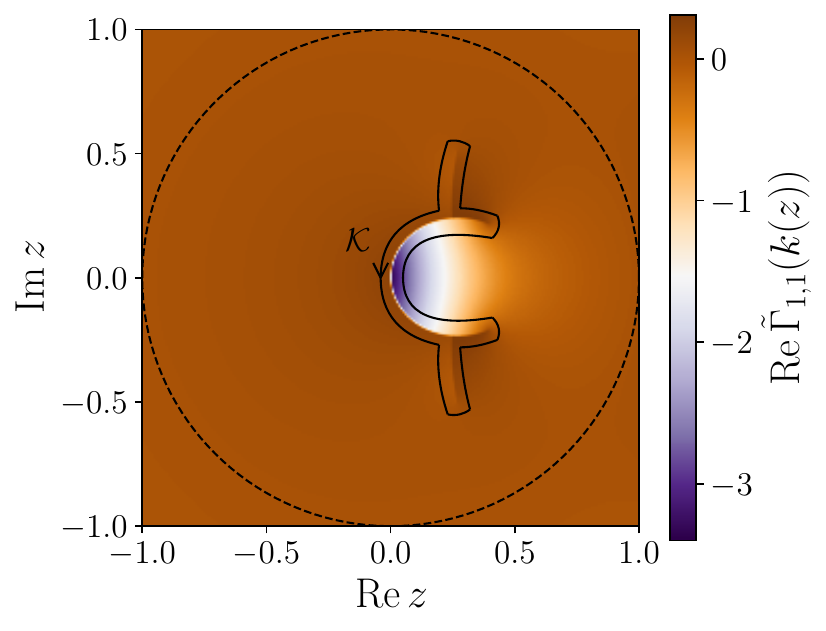}
 \caption{Correlation function $\Re \Tilde{\Gamma}_{1,1}$ after the change of variable $z=e^{i k}$, using $\mu = 0.4$, $\gamma=1$, and $q=0.5$. A sketch of the integration contour $\mathcal{K}$, encircling the discontinuity lines, is presented.}
 \label{f:circle_landscape}
\end{figure}

In this Appendix we show how to write the correlation matrix in terms of a contour integral, and how to bound it as in Eq.~\eqref{corr_bound}. From the definition of $\Tilde{\Gamma}(k)$, the inverse Fourier transform is given by
\begin{equation}
    \Gamma(x) = \frac{1}{L}\sum_k e^{i k x} \Tilde{\Gamma}(k),
\end{equation}
where the sum runs over the Brillouin zone given by $k=\pm\frac{2m-1}{L}\pi$, $m=1,\dots,L$, for periodic boundary conditions. For $L\to\infty$ the previous expression becomes an integral
\begin{equation}
    \Gamma(x) = \frac{1}{2\pi}\int_{-\pi}^\pi e^{i k x} \Tilde{\Gamma}(k) dk.
\end{equation}
After introducing the complex variable $z=e^{i k}$ we obtain the contour integral
\begin{equation}
    \Gamma(x) = \frac{1}{2\pi i}\oint_\mathcal{S} z^{x-1}\Tilde{\Gamma}(k(z)) dz,
\end{equation}
where $\mathcal{S}$ is the unit circle with counterclockwise orientation. This rewriting is possible assuming that $\Tilde{\Gamma}(k)$ is analytic for $k\in\mathbb{R}$, which is the case of non-critical steady states. In all regions where $\Tilde{\Gamma}(k(z))$ is regular the contour can be safely deformed. In particular, we can shrink $\mathcal{S}$ to a new contour $\mathcal{K}$ that encircles tightly the non-analiticities of the correlation matrix. This is represented in Fig.~\ref{f:circle_landscape}. With the introduction of $\mathcal{K}$ we finally obtain
\begin{equation}
    \Gamma(x) = \frac{1}{2\pi i}\oint_\mathcal{K} z^{x-1}\Tilde{\Gamma}(k(z)) dz,
\end{equation}

At last, we derive the upper bound to the correlation length. Let us parameterize the closed path $\mathcal{K}$ by a function $\kappa(s)$ of real variable $s\in [0,1)$, so that
\begin{equation}
    \Gamma(x) = \frac{1}{2\pi i}\int_{0}^1 \kappa(s)^{x-1}\Tilde{\Gamma}(k(\kappa(s))) \kappa'(s) ds.
\end{equation}
We then have
\begin{equation}
\begin{split}
    &\left|\Gamma(x)\right| \leq \frac{1}{2\pi}\int_{0}^1 \left|\kappa(s)\right|^{x-1}\left|\Tilde{\Gamma}(k(\kappa(s))) \kappa'(s)\right| ds\\
    &\leq \frac{1}{2\pi}\left(\max_{s\in [0,1)}\left\{\left|\kappa(s)\right|\right\}\right)^{x-1} \int_{0}^1\left|\Tilde{\Gamma}(k(\kappa(s))) \kappa'(s)\right| ds.
    \end{split}
\end{equation}
Since we take the contour $\mathcal{K}$ to be at an infinitesimal distance from the discontinuities of $\Tilde{\Gamma}$, the maximum of $\left|\kappa(s)\right|$ is achieved at the point of non-analiticity that is further away from the origin. Recalling that $z=e^{i k}$, the maximum modulus of $z$ corresponds to the minimum imaginary part of $k$, and thus
\begin{equation}
    \max_{s\in [0,1)}\left\{\left|\kappa(s)\right|\right\} = \exp\left(-\min_{k\in\mathcal{C}} \left\{\Im k\right\}\right).
\end{equation}
This yields the bound of Eq.~\eqref{corr_bound}.

\section{Third quantization}\label{a:third_quantization}
We now describe how to generalize the formalism of third quantization introduced in Refs.~\cite{prosen2008, prosen2010} to diagonalize Eq.~\eqref{liouv}. From now on we drop the constant term needed for trace preservation because it simply adds a shift to the Liouvillian spectrum, and we thus consider
\begin{equation}\label{reduced_liouv}
    \mathcal{L}\:\bullet = -i\comm{\hat{H}}{\bullet} + \sum_{j=1}^L\left((1-q)\hat{L}_j\bullet\hat{L}\daga_j-\frac{1}{2}\acomm{\hat{L}\daga_j\hat{L}_j}{\bullet}\right).
\end{equation}
The procedure is completely analogous to the one presented in Ref.~\cite{prosen2008}, and thus we will cover it only briefly.

The space of operators acting on the $2^N$-dimensional Hilbert space can be seen as a $4^N$-dimensional Hilbert space itself, whose elements $\hat{A}$ will be relabeled as $\ket{\hat{A}}$, when supplemented with the inner product $\braket{\hat{A}}{\hat{B}} = \Tr \left(\hat{A}\daga \hat{B}\right)/4^N$. The set of strings of Majorana operators 
\begin{equation}
    \hat{P}_{\boldsymbol{\alpha}} = \hat{w}_{1}^{\alpha_1}\hat{w}_{2}^{\alpha_2}\,...\, \hat{w}_{2N}^{\alpha_{2N}},
\end{equation}
where $\boldsymbol{\alpha} = \left(\alpha_1,\dots,\alpha_{2N}\right)$ and $\alpha_j=0,1$, forms a basis of the space of operators, and the states $\ket{\hat{P}_{\boldsymbol{\alpha}}}$ can be thought as Fock states. We introduce a set of annihilation and creation linear operators $\hat{f}_{j}$ and $\hat{f}_{j}\daga$, $j=1,\dots,2N$, defined by
\begin{subequations}
    \begin{equation}
        \hat{f}_j \ket{\hat{P}_{\boldsymbol{\alpha}}} = \delta_{\alpha_j,1}\ket{\hat{w}_j\hat{P}_{\boldsymbol{\alpha}}},
    \end{equation}
    \begin{equation}
        \hat{f}\daga_j \ket{\hat{P}_{\boldsymbol{\alpha}}} = \delta_{\alpha_j,0}\ket{\hat{w}_j\hat{P}_{\boldsymbol{\alpha}}}.
    \end{equation}
\end{subequations}
It is easily checked that $\hat{f}_j$ and $\hat{f}\daga_j$ fulfill canonical anticommutation relations. For convenience, let us introduce the $4 N$ Majorana operators $\hat{a}_{2j-1}= \left(\hat{f}_j + \hat{f}_j\daga\right)/\sqrt{2}$, $\hat{a}_{2j}= i\left(\hat{f}_j - \hat{f}_j\daga\right)/\sqrt{2}$. We remark that these operators are actually super-operators with respect to the original Hilbert space of physical states. The Liouvillian $\mathcal{L}$ can be represented as a sum of strings of $\hat{a}$ operators. In practice, this is done by evaluating how it acts on the basis states $\ket{\hat{P}_{\boldsymbol{\alpha}}}$. For the model of Eq.~\eqref{reduced_liouv} we obtain
\begin{equation}
\mathcal{L} = \sum_{m,n=1}^{2 N}\mathbb{A}_{m,n}\hat{a}_m \hat{a}_n - A_0
\end{equation}
where
\begin{equation}
    \mathbb{A} = \begin{pmatrix}
        -2i\mathbb{H} + 2i\Im M & 2i\left(1-q\right)M \\
        -2i\left(1-q\right)M^T & -2i\mathbb{H} - 2i\Im M
    \end{pmatrix},
\end{equation}
$\mathbb{H}$ and $M$ are the Hamiltonian and the bath matrices introduced in the main text in Eqs.~\eqref{hamiltonian_majorana} and~\eqref{bath_matrix}, and $A_0 = 2 \Tr M$.

Assuming that $\mathbb{A}$ is diagonalizable, it can be decomposed as
\begin{equation}
    \mathbb{A} = V^T \begin{pmatrix}
        0 & \beta_1 & 0 & 0 & \dots\\
        -\beta_1 & 0 & 0 & 0 & \dots\\
        0 & 0 & 0 & \beta_2 & \dots\\
        0 & 0 & -\beta_2 & 0 & \dots\\
        \vdots & \vdots & \vdots & \vdots & \ddots\\
    \end{pmatrix} V,
\end{equation}
where
\begin{equation}
    V V^T = \begin{pmatrix}
        0 & 1 & 0 & 0 & \dots\\
        1 & 0 & 0 & 0 & \dots\\
        0 & 0 & 0 & 1 & \dots\\
        0 & 0 & 1 & 0 & \dots\\
        \vdots & \vdots & \vdots & \vdots & \ddots\\
    \end{pmatrix}.
\end{equation}
The $\beta_m$ are called rapidities, and we can sort them in such a way that $\Re \beta_1 \geq \Re \beta_2 \geq \ldots \geq \Re \beta_{2 N} \geq 0$. This allows to introduce a set of normal master modes $\hat{b}_m = \sum_{n=1}^{4 N} V_{2m-1,n} \hat{a}_n$, $\hat{b}_m' = \sum_{n=1}^{4 N} V_{2m,n} \hat{a}_n$ that satisfy the almost-canonical anti-commutation relations $\acomm{\hat{b}_m}{\hat{b}_n} = \acomm{\hat{b}_m'}{\hat{b}_n'} = 0$, $\acomm{\hat{b}_m}{\hat{b}_n'}=\delta_{m,n}$. The Liouvillian can then be rewritten as
\begin{equation}
\mathcal{L} = -2\sum_{j=1}^{2 N} \beta_{j} \hat{b}_j' \hat{b}_j - B_0, \label{C1}
\end{equation}
where $B_0 = A_0 - \sum_{j=1}^{2 N} \beta_j$. The non-equilibrium steady state corresponds to the Liouvillian eigenoperator with the largest real part, and it thus corresponds to the vacuum state of the normal modes satisfying $\mathcal{L} \hat{\rho}_\text{NESS} = - B_0 \hat{\rho}_\text{NESS}$. It also follows, as in Ref.~\cite{prosen2008}, that the Liouvillian gap is given by $\Delta = -2 \operatorname{Re} \beta_{2 N}$.

\bibliography{biblio}

\end{document}